# Ordered state of magnetic charge in the pseudo-gap phase of a cuprate superconductor
### ($HgBa_2CuO_{4+\delta}$)

S. W. Lovesey[1,2] and D. D. Khalyavin[1]

[1]ISIS Facility, STFC Oxfordshire OX11 0QX, UK

[2]Diamond Light Source Ltd, Oxfordshire OX11 0DE, UK

**Abstract** A symmetry-based interpretation of published experimental results demonstrates that the pseudo-gap phase of underdoped $HgBa_2CuO_{4+\delta}$ (Hg1201) possesses an ordered state of magnetic charge epitomized by Cu magnetic monopoles. Magnetic properties of one-layer Hg1201 and two-layer $YBa_2Cu_3O_{6+x}$ (YBCO) cuprates have much in common, because their pseudo-gap phases possess the same magnetic space-group, e.g., both underdoped cuprates allow the magneto-electric (Kerr) effect. Differences in their properties stem from different Cu site symmetries, leaving Cu magnetic monopoles forbidden in YBCO. Resonant x-ray Bragg diffraction experiments can complement the wealth of information available from neutron diffraction experiments on five Hg1201 samples on which our findings are based. In the case of Hg1201 emergence of the pseudo-gap phase, with time-reversal violation, is accompanied by a reduction of Cu site symmetry that includes loss of a centre of inversion symmetry. In consequence, parity-odd x-ray absorption events herald the onset of the enigmatic phase, and we predict dependence of corresponding Bragg spots on magneto-electric multipoles, including the monopole, and the azimuthal angle (crystal rotation about the Bragg wavevector).

---

There is general agreement within the community of researchers that investigate high-$T_c$ superconducting materials that it is most important to fully understand the enigmatic pseudo-gap phase [1]. The phase exists below a temperature $T^*$ for less than optimal doping of holes at which the temperature of the superconducting transition is a maximum. No clear jump in the specific heat or change in lattice symmetry is observed on crossing $T^*$. However, in complex, inhomogeneous materials a phase transition is often diffuse and the entropy release is very difficult to isolate, e.g., with oxygen vacancy ordering.

Neutron Bragg diffraction experiments have demonstrated that underdoped $HgBa_2CuO_{4+\delta}$ (Hg1201) samples possess long-range magnetic order indexed on the chemical structure, and the same is found in identical experiments on a structurally more complicated cuprate $YBa_2Cu_3O_{6+x}$ (YBCO), with two $CuO_2$ plaquettes in a unit cell [2, 3]. In this communication we demonstrate that violation of time-reversal in Hg1201 on crossing $T^*$ is associated with the formation of Cu magneto-electric multipoles. A ferro-type motif of quadrupoles is found to be completely consistent with all available neutron diffraction data. The story for Hg1201 is similar to YBCO, where neutron magnetic Bragg diffraction and the Kerr effect have been observed and successfully related to ordered magneto-electric quadrupoles [2, 4, 5]. While Hg1201 and YBCO possess the same magnetic space-group (Cm'm'm'), not surprisingly, there are significant differences in the symmetries at sites used by Cu ions.

Symmetry in Hg1201 is even more restrictive than in YBCO. For, Cu ions are at specific and not general sites in the unit cell, while site symmetry forbids conventional (parity-even) magnetism and all dipoles. Thus, Hg1201 bears an ordered state of magnetic charge, revealed in a ferro-type motif of magneto-electric multipoles that deflect neutrons and x-rays. In addition, we predict that the Kerr effect is allowed in underdoped Hg1201.

The Cu site in the parent structure P4/mmm of Hg1201 is 1b at position (0, 0, 1/2), and the chemical structure is illustrated in Figure 1 [6]. The symmetry of the Cu site is 4/mmm ($D_{4h}$). The centre of inversion symmetry in 4/mmm forbids formation of parity-odd multipoles. A magnetic motif based on one Cu ion per cell must lead to Bragg spots indexed on the chemical structure and ferromagnetism, which has not been detected in the pseudo-gap phase of Hg1201. Construction of the correct magnetic space-group is initiated by P4/mmm1' and the absence of a ferromagnetism. One finds that the magnetic space-group Cm'm'm' is allowed. Copper ions use sites 2d at (0, 0, 1/2) with symmetry m'm'm', and a basis {(1, −1, 0), (1, 1, 0), (0, 0, 1)} with respect to the tetragonal parent is labelled (x, y, z). We find that this magnetic structure fits all available magnetic neutron diffraction data on underdoped Hg1201. Thereafter, we predict behaviour of resonant x-ray Bragg diffraction, because the experiments can bolster evidence already available from neutron diffraction. First, though, we note that Cm'm'm' belongs to the magnetic crystal-class m'm'm' that allows the Kerr effect [4].

The magnetic neutron diffraction amplitude for a scattering wavevector **k** can be derived from [7],

$$\langle \mathbf{Q} \rangle = \sum_\mathbf{d} \exp(i\mathbf{d} \cdot \mathbf{k}) \langle \exp(i\mathbf{R} \cdot \mathbf{k}) [\mathbf{S} - i(\mathbf{k} \times \mathbf{p})/\hbar k^2] \rangle_\mathbf{d}, \qquad (1)$$

where **R**, **S** and **p** are electron position, spin and linear momentum operators, respectively, and the sum **d** is over all positions occupied by magnetic ions in a unit cell. The scattering amplitude is $\langle \mathbf{Q}_\perp \rangle = [\boldsymbol{\kappa} \times (\langle \mathbf{Q} \rangle \times \boldsymbol{\kappa})]$, using a unit vector $\boldsymbol{\kappa} = \mathbf{k}/k$, and intensity of a Bragg spot is proportional to $|\langle \mathbf{Q}_\perp \rangle|^2$ in a conventional, unpolarized neutron diffraction experiment. In these expressions, angular brackets $\langle \ldots \rangle$ denote the expectation value, or time-average, of the enclosed operator.

The amplitude $\langle \mathbf{Q}_\perp \rangle$ is a sum of electronic multipoles $\langle U^K_Q \rangle$ where K is the rank and (2K + 1) projections obey − K ≤ Q ≤ K [7, 8]. Multipoles are defined to have definite discrete symmetries. The complex conjugate $\langle U^K_Q \rangle^* = (-1)^Q \langle U^K_{-Q} \rangle$, and our phase convention for real and imaginary parts of a multipole is $\langle U^K_Q \rangle = \langle U^K_Q \rangle' + i \langle U^K_Q \rangle''$.

Site symmetry m'm'm' requires,

$$\langle U^K_Q \rangle = I\, \theta\, C_{2y} \langle U^K_Q \rangle = \sigma_\pi \sigma_\theta (-1)^{K+Q} \langle U^K_{-Q} \rangle, \qquad (2)$$

where I, θ and $C_{2y}$ are operators for inversion, time reversal and rotation by 180° about the y-axis. In the second equality, $\sigma_\pi = \pm 1$ and $\sigma_\theta = \pm 1$ are parity and time signatures of $\langle U^K_Q \rangle$. However, the identity $\sigma_\pi \sigma_\theta = +1$ in m'm'm' means multipoles are either parity-even & time-even (electric charge) or parity-odd & time-odd (magnetic charge). A diad axis of symmetry on the c-axis restricts projections Q to even integers, Q = 0, ±2, ... Thus (2) reduces to,

$$\langle U^K_Q \rangle = (-1)^K \langle U^K_{-Q} \rangle = (-1)^K \langle U^K_Q \rangle^*. \qquad (3)$$

Evidently, $\langle U^K_0 \rangle = 0$ for K odd, and dipoles (K = 1) are forbidden.

Miller indices for orthorhombic Cm'm'm' are denoted (h, k, l), with $\kappa_x \propto h$, $\kappa_y \propto k$, and $\kappa_z \propto l$. For the parent tetragonal structure, P4/mmm, Miller indices are denoted ($H_o$, $K_o$, $L_o$), and h + k = $2H_o$, h − k = − $2K_o$ and l = $L_o$. The extinction rule h + k even is imposed by c-centring. Components of the scattering amplitude are,

$$\langle Q_a \rangle = (1/\sqrt{2})(\langle Q_y \rangle + \langle Q_x \rangle), \quad \langle Q_b \rangle = (1/\sqrt{2})(\langle Q_y \rangle - \langle Q_x \rangle), \quad \langle Q_c \rangle = \langle Q_z \rangle. \qquad (4)$$

Multipoles engaged in magnetic neutron scattering are time-odd, and thus parity-odd in the magnetic structure that we propose for Hg1201. Using (3) we find [7],

$$\langle Q_x \rangle \approx \kappa_x C [\langle H^2_0 \rangle - \sqrt{6} \langle H^2_{+2} \rangle'], \quad \langle Q_y \rangle = \kappa_y C [\langle H^2_0 \rangle + \sqrt{6} \langle H^2_{+2} \rangle'], \qquad (5)$$

$$\langle Q_z \rangle \approx - 2 \kappa_z C \langle H^2_0 \rangle,$$

with C = $-i\sqrt{(6/5)} (-1)^{L_o}$. We omit multipoles of higher rank than quadrupoles in (5), because their contributions are likely very small [7]. Magneto-electric quadrupoles in (5) are $\langle H^2_0 \rangle \propto \langle 3S_z n_z - \mathbf{S} \cdot \mathbf{n} \rangle$ and $\langle H^2_{+2} \rangle' \propto \langle S_x n_x - S_y n_y \rangle$ where **n** is the electric dipole operator. These parity-odd multipoles probe hybridized states 3d(Cu)-4p(Cu) and 3d(Cu)-2p(O), and model calculations are reported in references [7, 10].

Bragg spots indexed (0, 0, $L_o$) have $\kappa_x = \kappa_y = 0$, and from (5) one finds $\langle \mathbf{Q} \rangle = (0, 0, \langle Q_z \rangle)$ with the result $\langle \mathbf{Q}_\perp \rangle = 0$. This result explains the observation by Li *et al* of zero intensity in the Bragg spot (0, 0, 3) [3].

Data using neutron polarization analysis have also been reported [3]. The technique measures the magnetic contribution to the intensity of a Bragg spot with overlapping nuclear and magnetic amplitudes, which occurs when the magnetic motif and chemical structure coincide [14]. Primary and secondary polarizations are **P** and **P'**, and a fraction (1 − **P** • **P'**)/2 of neutrons participate in events that change (flip) the neutron spin orientation. For a collinear magnetic motif one finds (1 − **P** • **P'**)/2 ∝ {(1/2) (1 + $P^2$) |$\langle \mathbf{Q}_\perp \rangle$|$^2$ − |**P** • $\langle \mathbf{Q}_\perp \rangle$|$^2$} [14, 15]. A quantity SF = {|$\langle \mathbf{Q}_\perp \rangle$|$^2$ − |**P** • $\langle \mathbf{Q}_\perp \rangle$|$^2$} obtained with $P^2$ = 1 is a convenient measure of the strength of spin-flip scattering. Data for SF are reported at various temperatures [3]. Five Hg1201 samples with different doping levels were examined, and magnetic intensity grew as the temperature was lowered through $T^*$ in all cases.

To relate our theory to reported observations we must use $\kappa_a = k_a/k \propto H_o$, $\kappa_b = k_b/k \propto K_o$, $\kappa_c = k_c/k \propto L_o$ to define a unit vector $\mathbf{k}/k = (\kappa_a, \kappa_b, \kappa_c)$. Three types of polarization analysis were used at the Bragg spot (1, 0, 1) for which $\mathbf{k}/k = (\kappa_a, 0, \kappa_c)$ with ($\kappa_a^2 + \kappa_c^2$) = 1, while $\kappa_x = \kappa_y$ and $\kappa_z/\kappa_x$ = (a$\sqrt{2}$/c) [3, 6]. We find,

$$\langle \mathbf{Q}_\perp \rangle = (-\kappa_c (\kappa_a \langle Q_c \rangle - \kappa_c \langle Q_a \rangle), \langle Q_b \rangle, \kappa_a (\kappa_a \langle Q_c \rangle - \kappa_c \langle Q_a \rangle)), \qquad (6)$$

that gives;

(a) **P** // **k**, SF(a) = |$\langle \mathbf{Q}_\perp \rangle$|$^2$.

(b) **P** • **k** = 0 using **P** = (−$\kappa_c$, 0, $\kappa_a$) and **P** • $\langle \mathbf{Q}_\perp \rangle$ = ($\kappa_a \langle Q_c \rangle - \kappa_c \langle Q_a \rangle$) leads to,

SF(b) = |$\langle Q_b \rangle$|$^2$.

(c) **P** • **k** = 0 using **P** = (0, 1, 0) and **P** • $\langle \mathbf{Q}_\perp \rangle$ = $\langle Q_b \rangle$ leads to,

$$SF(c) = |\kappa_a \langle Q_c \rangle - \kappa_c \langle Q_a \rangle|^2. \tag{7}$$

(Labelling (a), (b) and (c) in (7) exactly follows the labelling of panels used to report data for SF, Figure 1 in reference [3].) Intensity SF(b) demonstrates that Cu ions possess magnetic multipoles with components in the basal plane, while SF(c) relates to the component of $\langle Q_\perp \rangle$ parallel to the crystal c-axis. The observation is that strengths of SF(b) and SF(c) are more or less equal [3]. From predictions in (5) we find $\langle Q_a \rangle \propto \kappa_x \langle H^2_0 \rangle$, $\langle Q_b \rangle \propto \kappa_x \langle H^2_{+2} \rangle'$, $\langle Q_c \rangle \propto \kappa_z \langle H^2_0 \rangle$, i.e., results (b) and (c) measure the individual components of the magneto-electric quadrupole with SF(b) $\propto (\langle H^2_{+2} \rangle')^2$ and SF(c) $\propto (\langle H^2_0 \rangle)^2$. The polarization sum-rule SF(a) = SF(b) + SF(c) is fulfilled by data gathered on Hg1201, which is paraded by the authors as further evidence of the magnetic nature of the diffraction signal [2, 3].

While magnetic dipole moments are forbidden in our theory of the pseudo-gap phase of underdoped Hg1201, nonetheless it is worth noting what they could imply for intensities of Bragg spots in magnetic neutron diffraction if they did exist. Null intensity in Bragg spots indexed $(0, 0, L_o)$ could mean moments have no components in the plane normal to the c-axis. Yet, SF(b) is different from zero in the experiments and would be proportional to the moment parallel to the b-axis in a conventional analysis. The conundrum exposed for magnetic moments has no place in our model, with its exclusive use of magneto-electric multipoles.

Bragg diffraction of x-rays enhanced by a Cu atomic resonance can add significant weight to our understanding of the pseudo-gap phase derived from neutron Bragg diffraction [8, 9, 10]. We begin with parity-even events at Cu L-edges, and thereafter parity-odd events at L-edges and the K-edge. States of photon polarization are defined in Figure 2, and our unit-cell structure factors include dependence on the rotation of the crystal through an angle $\psi$ around the Bragg wavevector in a so-called azimuthal-angle scan.

Parity-even multipoles observed in diffraction using an E1-E1 absorption event, at the Cu $L_2$ (0.965 keV) and $L_3$ (0.945 keV) absorption edges, probe 3d(Cu) states. We denote corresponding multipoles by $\langle T^K_Q \rangle$, and their time signature is $(-1)^K$ [8, 10]. Since multipoles allowed in the pseudo-gap phase possess $\sigma_\pi \sigma_\theta = +1$ it follows that $\sigma_\pi = +1$ and $\sigma_\theta = +1$ with K even. Convenient for the material under investigation are Bragg reflections $(0, 0, L_o)$, and the rotated channel of polarization $\pi'\sigma$ is always particularly useful. (Unit-cell structure factors for other Bragg reflections and unrotated channels of polarization are readily derived from universal expressions published by Scagnoli and Lovesey [9]). The corresponding unit-cell structure factor $F_{\pi'\sigma}$(E1-E1) is a function of a quadrupole $\langle T^2_{+2} \rangle$ that is purely real. We find,

$$F_{\pi'\sigma}(E1\text{-}E1) = - F_{\sigma'\pi}(E1\text{-}E1) = - \sin(\theta) \sin(2\psi) \, \langle T^2_{+2} \rangle'. \tag{8}$$

A factor $\{2(-1)^{L_o}\}$ is omitted from (8), and both subsequent unit-cell structure factors. Intensity of the Bragg spot $|F_{\pi'\sigma}(E1\text{-}E1)|^2$ is zero at the origin of the azimuthal-angle scan ($\psi = 0$), and the harmonic dependence $\sin(2\psi)$ is a direct consequence of the diad axis of symmetry in the Cu site. A tetrad axis of symmetry prevails in the parent structure. Referring to Figure 2, the basis vector $(1, -1, 0)$ is parallel to $-z$ at $\psi = 0$, and the crystal c-axis is aligned with $-x$.

Magneto-electric multipoles ($\sigma_\pi = -1$ and $\sigma_\theta = -1$) exist because time-reversal is violated, and therefore appear alongside magnetic Bragg spots in neutron diffraction patterns as evidence of magnetic order. We denote magneto-electric multipoles by $\langle G^K_Q \rangle$ [8, 9, 10]. The scalar $\langle G^0_0 \rangle \propto \langle S \cdot n \rangle$ is a magnetic charge (monopole) that has been revealed in simulations of lithium orthophosphates [11].

An interpretation of resonant x-ray Bragg diffraction by antiferromagnetic CuO appealed to an E1-M1 absorption event at Cu L-edges [12, 13]. Intensity of Bragg spots indexed by $(0, 0, L_o)$ for Hg1201 enhanced by an E1-M1 event are related to a structure factor,

$$F_{\pi'\sigma}(E1\text{-}M1) = F_{\sigma'\pi}(E1\text{-}M1) = -(2/\sqrt{3})\sin^2(\theta)\langle G^0_0\rangle$$

$$+ (1/\sqrt{6})[2 + \cos^2(\theta)]\langle G^2_0\rangle - \cos^2(\theta)\cos(2\psi)\langle G^2_{+2}\rangle'. \qquad (9)$$

Contributions to structure factors from diagonal multipoles $\langle G^K_0\rangle$ are understandably independent of the azimuthal angle. Azimuthal-angle dependence of E1-E1 and E1-M1 enhanced diffraction amplitudes differ by 45°. Using E = 0.945 keV for the photon energy we find the reflection (0, 0, 1) falls in the Ewald sphere with $\sin(\theta) = 0.692$ [6].

Enhancement by an E1-E2 event might be found in the vicinity of the Cu K-edge (8.993 keV) and probes 3d(Cu), 4p(Cu) and 2p(O) states. The corresponding unit-cell structure factor can contain multipoles with ranks K = 1, 2, 3 and for Bragg spots $(0, 0, L_o)$ in diffraction from Hg1201 we find,

$$F_{\pi'\sigma}(E1\text{-}E2) = F_{\sigma'\pi}(E1\text{-}E2) = (1/2\sqrt{5})[2 - 3\cos^2(\theta)]\langle G^2_0\rangle$$

$$+ (1/\sqrt{30})\cos^2(\theta)\cos(2\psi)[\langle G^2_{+2}\rangle' - 2\sqrt{2}\langle G^3_{+2}\rangle'']. \qquad (10)$$

Note that unit-cell structure factors (8), (9) or (10) are purely real, unlike the neutron scattering amplitude that is purely imaginary. Azimuthal-angle dependence in E1-M1 and E1-E2 amplitudes are the same but they differ with respect to dependence on the Bragg angle, $\theta$.

Observation of x-ray diffraction Bragg spots in accord with results (8), (9) or (10) will add yet more credibility to our model of the pseudo-gap phase of Hg1201 - that already accounts for the available magnetic neutron diffraction data.

To conclude, we find that extensive sets of magnetic neutron diffraction data gathered on five underdoped Hg1201 samples fit the magnetic space-group Cm'm'm' [2, 3], which allows the Kerr effect [4]. The same space group applies to like diffraction data for the pseudo-gap phase of many YBCO samples [5], a result that underpins a universality of underdoped YBCO and Hg1201 properties highlighted by Li *et al* [3]. Although underdoped Hg1201 possesses Cu magnetic monopoles unlike YBCO, according to our theory. Moreover, an ordered state of pure magnetic charge exists in the pseudo-gap phase of Hg1201. We propose resonant x-ray diffraction experiments on Hg1201 to validate specific features of our theoretical understanding of the pseudo-gap phase of Hg1201 and its emergence from the parent state.

**Acknowledgement** We are grateful to Dr Urs Staub for discussions and correspondence about magnetic neutron diffraction by cuprates. Professor Ewald Balcar prepared Figure 2, and helped SWL with ongoing scrutiny.

**Figure 1** Chemical structure of $HgBa_2CuO_{4+\delta}$ (Hg1201) with cell dimensions a ≈ 3.885 Å and c ≈ 9.485 Å [6]. There is one $CuO_2$ plaquette per unit cell, and Cu is at the site 1b (0, 0, 1/2) with 4/mmm ($D_{4h}$) symmetry. A Cu ion is at a centre of inversion symmetry.

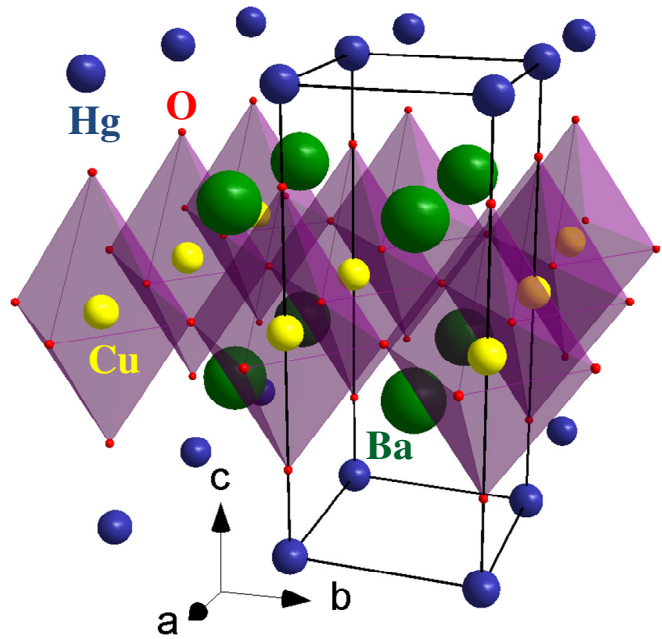

**Figure 2** The diagram illustrates the Cartesian coordinate system (x, y, z) adopted for resonant Bragg diffraction of x-rays and the relation to states of polarization, σ and π, in the primary (unprimed) and secondary (primed) beams. In the nominal setting of the crystal the system (x, y, z) coincides with basis vectors {(1, −1, 0), (1, 1, 0), (0, 0, 1)}, which are also given the same Cartesian labels. The beam is deflected through and angle 2θ, and **q** and **q'** are primary and secondary wavevectors. A Bragg wavevector **q** − **q'** = τ(h k l) and for τ(h k l) = (0, 0, $L_o$) in cases (8), (9) & (10) the crystal c-axis is parallel to −x, as indicated in the diagram. At the origin of an azimuthal-angle scan (ψ = 0) the basis vector (1, −1, 0) is aligned with −z.

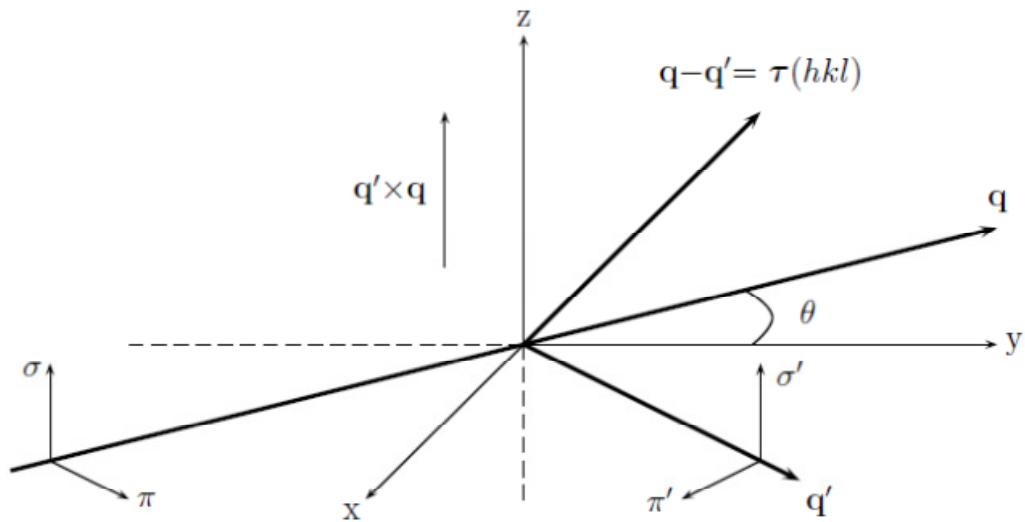